\documentclass[10pt,conference,letterpaper]{IEEEtran}
\usepackage{cite}
\usepackage{graphicx}
\usepackage{float}
\usepackage{amsmath}
\usepackage[utf8]{inputenc}
\usepackage[final]{pdfpages}
\usepackage{pdfpages}
\usepackage{lipsum}
\usepackage{textcase}
\usepackage{url}
\usepackage{times,amsmath,epsfig}
\usepackage{epstopdf}
\usepackage{array}
\usepackage{amssymb}
\usepackage{subfigure}
\usepackage{subfig}
\usepackage{etoolbox}% http://ctan.org/pkg/etoolbox
\AtEndEnvironment{remark}{\null\hfill\qedsymbol}%

\AtEndEnvironment{definition}{\null\hfill\qedsymbol}%

	\newcommand\blfootnote[1]{%
		\begingroup
		\renewcommand\thefootnote{}\footnote{#1}%
		\addtocounter{footnote}{-1}%
		\endgroup
	}
\newcolumntype{P}[1]{>{\centering\arraybackslash}p{#1}}
\newcolumntype{M}[1]{>{\centering\arraybackslash}m{#1}}

\begin{document}
	
\title{Micro-UAV Detection  with a Low-Grazing Angle Millimeter Wave Radar}
\author{%
\normalsize Martins Ezuma, Ozgur Ozdemir, Chethan Kumar Anjinappa, Wahab Ali Gulzar  and Ismail Guvenc\\
Department of Electrical and Computer Engineering, North Carolina State University, Raleigh, NC,\\
Email: \{mcezuma, oozdemi, canjina, wkhawaj, iguvenc\}@ncsu.edu}

\maketitle
\blfootnote{This work has been supported by the NASA grant NNX17AJ94A.}
% * <mcezuma@ncsu.edu> 2018-06-16T23:53:31.005Z:
%
% ^.
\begin{abstract}
Millimeter wave radars are popularly used in last-mile radar-based defense systems. Detection of low-altitude airborne target using these radars at low-grazing angles is an important problem in the field of electronic warfare, which becomes challenging due to the significant effects of clutters in the terrain. This paper provides both experimental and analytical investigation of micro unmanned aerial vehicle (UAV) detection in a rocky terrain using a low grazing angle, surface-sited 24 GHz dual polarized frequency modulated continuous wave (FMCW) radar. The radar backscatter signal from the UAV is polluted by land clutters which is modeled using a uniform Weibull distribution. A constant false alarm rate (CFAR) detector which employs adaptive thresholding is designed to detect the UAV in the rich clutter background. In order to further enhance the discrimination of the UAV from the clutter, the micro-Doppler signature of the rotating propellers and bulk trajectory of the UAV are extracted and plotted in the time-frequency domain.    


\end{abstract}

\IEEEpeerreviewmaketitle

\section{Introduction}
% According to a recent sales forecast by the United States Federal Aviation Administration (FAA), by 2020, there will be about seven million new micro-Unmanned Aerial Vehicles (UAVs, popularly known as Drones) flying over the the U.S airspace~\cite{FCC-report}.The report estimated that hobbyist will accounting for more than 60\% of this number. However, in recent times, there have been occurrence of civilian drones crashing into aircrafts and many drone users have been accused of intentionally violating no-fly zones~\cite{FCC-report1}. Therefore, there is an urgent need to effectively detect, track and classify micro-UAVs on flight...

Low-grazing angle targets like micro unmanned aerial vehicles (micro-UAVs) are difficult to detect using radars, making them a potential air defense threat. Traditional radar detection techniques using low frequency pulse radars are not effective in detecting micro-UAVs, since these targets have very small radar cross section (RCS) and they fly at low attitudes~\cite{FCC_report2}. Recently, millimeter wave (mmWave) radar has been studied as a potential solution for the detection/tracking of micro-UAVs. This is because the high resolution property of these radars make them well suited for extracting the micro-Doppler signatures associated with UAVs. However, experimental and analytical studies have shown that at low grazing angles, the probability of detection of mmWave radars is %greatly 
 degraded by clutters from the terrain~\cite{rosenberg2015radar,ding2013low, jay2000evaluation}. 
 
In order to accurately detect a low-altitude micro-UAV, the clutter statistics have to be considered. Experimental studies of radar clutters have long established that these clutters are non-Rayleigh, non-stationary and at best can be described by distributions such as Weibull, Pareto, and K-distributions~\cite{billingsley1993ground}. These non-Rayleigh clutter/noise model will require the development of an adaptive threshold detector for target detection and examples of such detectors are the constant false alarm rate (CFAR) detector, generalized likelihood ratio test-linear quadratic (GLRT-LQ) detector and the Bayesian optimum radar detector (BORD). The latter is optimal when the clutter is modeled using a spherically invariant random process (SIRP) model.
  
%Although, in recent times, 
While several researchers have recently investigated the problem of micro-UAV detection using mmWave radars~\cite{drozdowicz201635, kim2018drone}, to the best of our knowledge, none of these works have considered the effects of the non-Rayleigh clutter statistics of the terrain. In this paper, we develop a cell-averaging CFAR (CA-CFAR) processor for automatic detection of a micro-UAV in rocky terrain. The land terrain clutter is modeled using the Weibull distribution. 

The rest of this paper is organized as follows. Section~\ref{Sec2} briefly describes the system; Section~\ref{Sec3} describes the automatic detection system, clutter model, and UAV micro-Doppler signature; Section~\ref{Sec:NumRes} provides numerical results and some concluding remarks. %, and finally the last section concludes the paper. 

%  The result of the CA-CFAR detector coupled with the spectrogram display of the extracted radar micro-Doppler signature provides sufficient information for the detection and discrimination of the UAV. Monte Carlos experimental simulation is used to evaluate the performance of the detector.

% \begin{figure}[!t]
%  \center
%  \includegraphics[width=85mm]{system_setup4.PNG}\vspace{-1mm}
%  \caption{Experimental setup with ground clutter reflections considering a grazing angle of $\psi$.}\vspace{-4mm}\label{Fig:systemsetup2}
% \end{figure}
% %\vspace{-3cm}

\section{System Model for UAV Detection} \label{Sec2}

\begin{figure}[t!]
%\vspace{-0.5cm}
 \center
 \includegraphics[width=85mm]{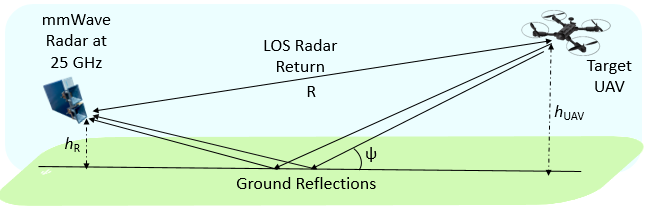}\vspace{-1mm}
 \caption{Drone detection with ground clutter reflections considering a grazing angle of $\psi$.}\vspace{-2mm}\label{Fig:system_setup5}
\vspace{-0.5cm}
\end{figure}

% \textcolor{red}{IG: You start with experiments but we have limited experiments. I would start more general, and mention experiments only when you describe them, to reduce emphasis. E.g. you can move the first paragraph to somewhere later on, and move Fig. 1 here, and mention that you consider the problem of UAV detection as illustrated in the figure, without referring to experiments. Section title can be ``system Model for UAV Detection''}

The problem of UAV detection considered herein is illustrated in Fig.~\ref{Fig:system_setup5}. The target is a low altitude micro-UAV, flying within the line of sight of the mmWave radar which sits a few meters above the earth. The radar receives a direct ray from the transmitter and indirect rays backscattered from the ground (clutter) at grazing angle, $\psi$. The clutter power represent an interference that degrades the target detection and this effect can be characterized by the signal-to-clutter ratio. Theoretically, the maximum detectable range of the radar is given by the equation:

% For the experiment, the K-band Ancortek 2400AD2 Software Defined Radar (SDR) was used. This is a high resolution dual polarized millimeter wave radar with center frequency at 25 GHz. The radar has three operational modes: FMCW, CW and FSK. For the application described herein, the FMCW mode is used with parameters specified in Table~\ref{Table:Table1}. 

%      The experimental setup is shown in Fig.~\ref{Fig:systemsetup2}. The target is a Phantom 4 Pro DJI UAV. This is a small UAV with a very low radar cross section (RCS) flying at a low altitude, making detection difficult for a low grazing angle radar.  Theoretically, we can simulate the detection range of the radar system by means of the radar range equation:

\begin{equation}
 \label{eq:1}
% R=\sum_{p=0}^{N_{\rm P}-1} a_p \exp(j\theta_p) \tilde{s}(n-\tau_p).
R=\sqrt[4]{\frac{P_t G^{2}\lambda^2\sigma}{P_r(4\pi)^3L}},
\end{equation}
where $R$ is the range to the target, $P_t$ and $P_r$ are radar transmit and receive power respectively, $G$ is antenna gain, $\lambda$ is wavelength, $\sigma$ is radar cross section of target, and $L$ is the total loss (propagation and system losses). Alternatively, in terms of the signal-to-clutter-ratio (SCR) and the radar grazing angle $\psi$, the detection range of a target in an area clutter is given in~\cite{rountree1990radar} as:
\begin{align}
 \label{eq:2}
 R=\frac{L\cos(\psi)\sigma_t}{S_{\rm SCR}(\frac{c\tau}{2})\theta\sigma_0} ~{\rm for}~\tan\psi<\frac{\phi R}{c\tau/2},
\end{align} 
% \vspace{-0.1cm}
where $R$ is the detection range, $L$ is the total  system loss, $\sigma_t$ is the target RCS, $S_{\rm SCR}$ is the SCR, $\sigma_0$ is the terrain backscatter (clutter) RCS per unit surface area ($m^2/m^2$), $\theta$ is 3-dB beamwidth in azimuth, 
$\phi$ is 3-dB beamwidth in elevation, $\tau$ is the transmitted pulse duration, and $c$ is the speed of light.

% where $N_{\rm P}$  
  
\begin{table}[t]
\centering
\caption{Ancortek FMCW Radar Parameters}
\label{Table:Table1}

\begin{tabular}{|l|c|c|c|}
\hline
Parameter & Unit & Value\\
\hline
Center Frequency & GHz & 24 \\
Bandwidth & GHz & 0.5 \\
Sweep Time & ms & 1\\
Number of Samples per Sweep &  & 128\\
Transmit Power& dBm & 12\\
Pulse repetition frequency (PRF) & KHz & 1\\
Range Resolution & m & 0.3\\
Antenna Gain & dBi & 20\\
Noise Figure over RX & dB & 6.4\\
% \hline
\hline
% Recommended & yes & yes & no & no & no\\
% \hline
\end{tabular}
\end{table}

% \begin{figure}[!h]
%  \center
%  \includegraphics[width=80mm]{Radar_System_Block_diagram.PNG}
%  \caption{Radar System}\label{Fig:systemsetup1}
% \end{figure}
% \textcolor{red}{IG: I changed the section title below to be more specific; it is also a subsection title but that is fine.}

\section{Constant False Alarm Detector}\label{Sec3}
 \subsection{Target Detection Problem Formulation}
\begin{figure}[!t]
\vspace{-.35cm}
\centering
 \includegraphics[scale=0.525]{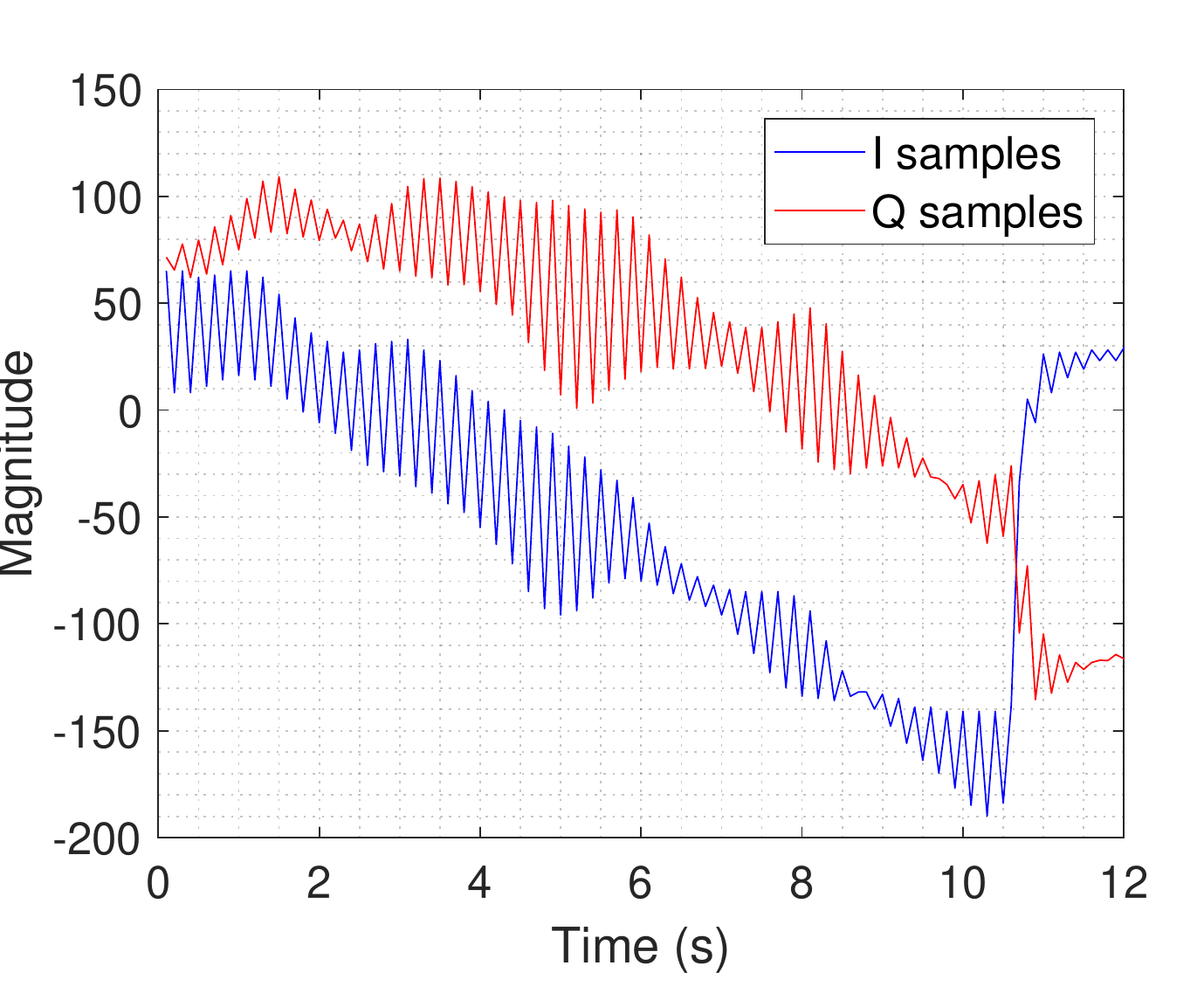}\vspace{-.3cm}
 \caption{Experimental radar I/Q data samples extracted from backscattered signal at the receiver of the 25 GHz Ancortek radar. The backscattered signals are reflected from both DJI phantom 4 Pro micro-UAV and the ground clutter.}\label{Fig:Radar IQ samples}\vspace{-.65cm}
\end{figure}

 Considering the system model in Fig.~\ref{Fig:system_setup5}, we are interested in the problem of detecting a complex signal $\textbf{s}$, backscattered from the UAV, in a set of $M$ radar measurement of complex vectors $\bf{y}=\bf{y}_{\rm I}+\bf{y}_{\rm Q}$, shown in Fig.~\ref{Fig:Radar IQ samples}. %, which have been 
This signal is corrupted by an independent additive interference, corresponding to the ground clutter reflections, \textbf{c}, and white Gaussian thermal noise at the receiver. However, for low altitude target, it is reasonable to assume that the interference caused by the ground clutter is more significant than the Gaussian noise, leaving the ground clutters as the overwhelming interference corrupting the received signal. For this model, the target detection problem is presented as a  hypothesis test as:
\begin{align}
\label{eq:3}
H_0:~\textbf{y} =  \textbf{c}, \qquad H_1:~\textbf{y}  =  \textbf{s}+\textbf{c}.
\end{align}
%\vspace{-0.25cm}
% where $H_0$ is the target-absent hypothesis (i.e. clutter plus noise only) and $H_1$ is a target present with clutter and noise hypothesis. The optimal decision is based on the Neyman-Pearson criteria, which optimizes the probability of detection ($P_{\rm d}$) while constraining the probability of false alarm ($P_{\rm fa}$). According to this criteria, the optimal target detection is given by the likelihood ratio test:
% \begin{align}
%  \label{eq:5}
% \Lambda(y) = \frac{P_y(y|H_1)}{P_y(y|H_0)}\underset{H_0}{\overset{H_1}{\gtrless}} T,
% \end{align}
% where $T$ is the detection threshold. However, at low grazing angle, the ground clutter power is continuously fluctuating and so we cannot set a fixed threshold, $T$, for the target detection. Therefore, there is a need to adaptively estimate the clutter power in order to maintain a constant $P_{\rm fa}$, while maximizing the probability of detection, $P_{\rm d}$.
Moreover, to discriminate the UAV from other potential aerial targets such as birds and aircrafts, we need to extract the unique micro-Doppler signature of its propellers. Furthermore, the trajectory of the UAV could also help us discriminate it from other aerial objects. 
%\vspace{-0.5cm}
% \begin{figure}[!t]
%  \center
%  \includegraphics[width=80mm]{Radar_I_Q_samples_fmcw21.pdf}
%  \caption{Radar I and Q data samples extracted from backscattered signals from UAV and clutter}\label{Fig:Radar IQ samples}
% \end{figure}
% \begin{equation}
%  \label{eq:5}
% \Lambda(y) = \frac{P_y(y|H_1)}{P_y(y|H_0)}\underset{H_0}{\overset{H_1}{\gtrless}} T
% \end{equation}
% where the optimal decision on whether the target is present or not is based on the Neyman-Pearson criterion which can be written as the likelihood ratio of the probability distribution of the target and target plus interference

\subsection{Weibull Clutter}
 Extensive experimental studies %at the MIT Lincoln laboratory 
 have shown that the amplitude (or power) of ground clutters illuminated at low-grazing angles by mmWave surfaces-sited radars can be accurately modeled by the Weibull distribution~\cite{billingsley1993ground}, which is characterized by two parameters: shape parameter ($k$) and scale parameter ($b$). In fact, the Weibull distribution is very flexible and can be fitted to different clutter types by changing its shape parameter. The cumulative distribution  function (CDF), the probability distribution function (PDF), and mean power (second moment) of the Weibull distribution are respectively given as follows: 
 \vspace{-0.2cm}
 \begin{align}
\label{eq:6}
F(z) &= 1-\exp\left[-\left(\frac{z}{b}\right)^k\right]~, \\ %\big]\Big]\bigg]\Bigg],  \\
\label{eq:7}
f(z)&=\frac{{\rm d} F(z)}{{\rm d}z}=\frac{k}{b}\left(\frac{z}{b}\right)^{k-1}\exp\left[-\left(\frac{z}{b}\right)^k\right]~,\\
\label{eq:7_1}
P&=b\Gamma\left(1+\frac{2}{k}\right), 
\end{align}
where $\Gamma(.)$ denotes the Gamma function.

% \cref can be used to reference range/multiple equations in latex % i did not use it here. to use it you need package cref package

\subsection{Constant False Alarm Detector (CFAR)}
 In order to adaptively detect the micro-UAV in the presence of significant ground clutters, the range-Doppler plot is created from the received I and Q data samples. Each pixel on the range-Doppler plot is called a range cell and represent the potential position of a point target (or scatterer). For a given pulse bandwidth $B$, the resolution or bin size of each range cell is $\Delta_{\rm res} \approx \frac{c}{2B}$. Therefore, a target is detected at a cell position if the target power is greater than the surrounding clutter power. 
 
 In the cell averaging CFAR (CA-CFAR), the adaptive threshold for the detection of a target in any given cell under test (CUT) is set by estimating the local mean clutter power of the neighboring cells (or pixels) by using a sliding window estimator. In each instance, a single CUT is centered in the sliding window with equal number of leading and lagging neighboring cells. The detection threshold for the CUT is given as 
%\begin{equation}
% \label{eq:9}
$T=\alpha \hat{P}$, 
%\end{equation}
where $\hat{P}$ is the estimated local mean clutter power and $\alpha$ is the threshold multiplying factor. In conventional CA-CFAR, $\alpha$ is defined in terms of the desired $P_{\rm fa}$ as $\alpha =N(P_{\rm fa}^{-1/N}-1)$, where $N$ is the total number of neighboring cells~\cite{richards2014fundamentals}. Moreover, if the length of the sliding window is very small compared to the width of the Range-Doppler plot, it is reasonable to assume that the local clutter power around a given CUT is homogeneous.

This homogeneous local clutter power distribution can be represented by $N$ independent and identically distributed (iid) Weibull random variables with average power given by  
%equation 
(\ref{eq:7_1}). Therefore, in order to estimate $P$, we need to estimate the values of the scale parameter $b$ for a given value of $k$ using parametric techniques like the maximum likelihood (ML), maximum entropy, and the method of moments (MOM). In~\cite{datsiou2018weibull}, the ML estimate of the scale parameter $b$ of a sample of $N$ independent and identically distributed Weibull random variables is given~as: 
%\vspace{-0.5cm} %IG: Never over-use the negative vspace, looks very ugly!
%\vspace{-0.5cm}
\begin{align}
 \label{eq:10}
\hat{b}_{\rm MLE}&=\frac{1}{N}\sum\limits_{i=1}^N (z_{i}^k)^\frac{1}{k}~,\\
% \end{align}
% \begin{align}
\label{eq:11}
F(z)&=1-\exp\left[-\left(\frac{T}{\hat{b}_{\rm MLE} }\right)^k\right]\underset{H_0}{\overset{H_1}{\gtrless}} 1-P_{\rm fa}~.
\end{align}

\begin{figure}[t!]
 \center
 \includegraphics[width=0.4\textwidth]{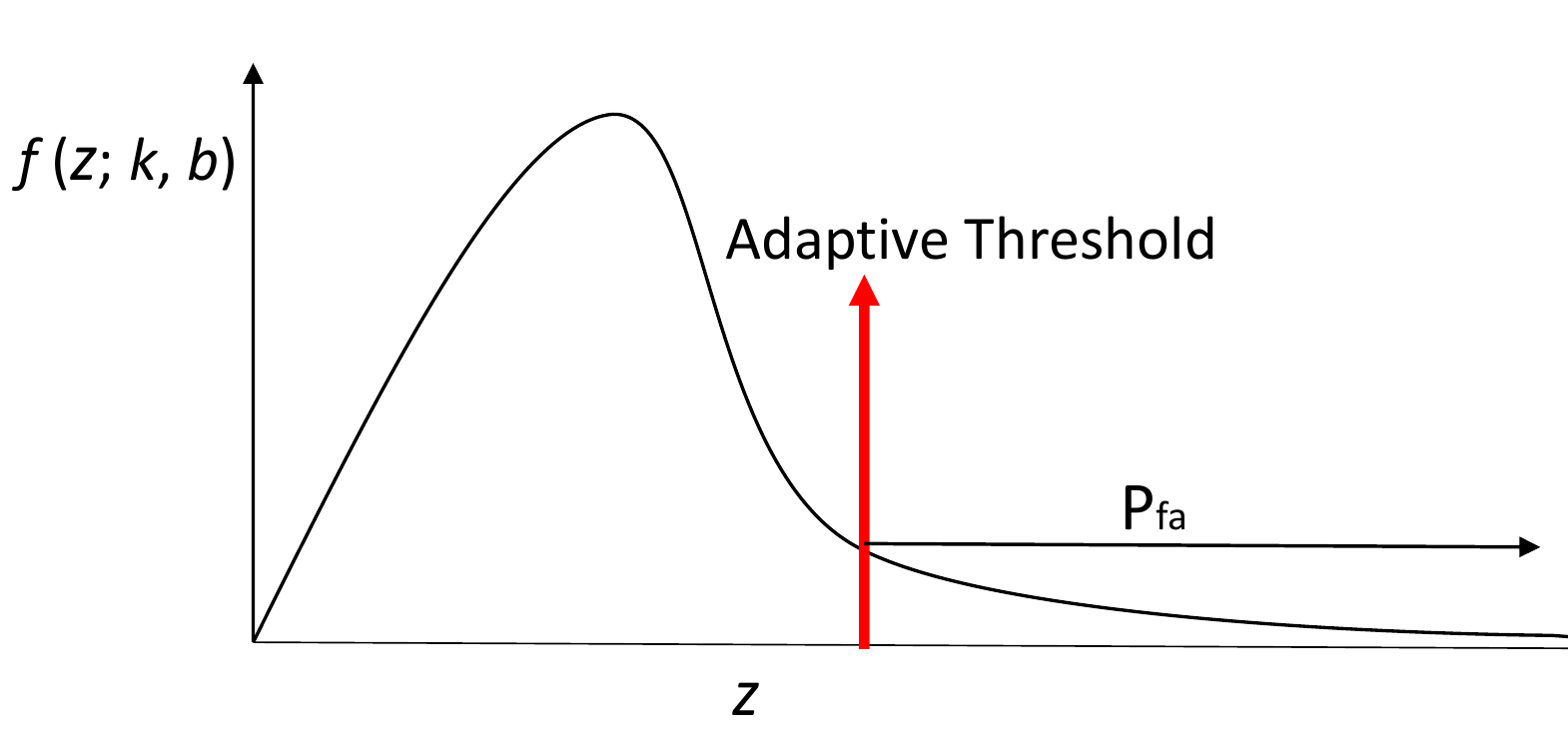}
 \caption{Adaptive CFAR threshold on the heavy tailed Weibull clutter distribution.}\label{Fig:Adaptivethreshold}
\end{figure}

Once the scale parameter of the Weibull distribution is estimated, the local clutter power and detection threshold can be estimated adaptively from (\ref{eq:7_1}) and $\alpha\hat{P}$, respectively. Therefore, the decision about the pixel (i.e., the CUT) can be made with the CFAR detector by means of the hypothesis test in (\ref{eq:11}). The optimal adaptive threshold for a given $P_{\rm fa}$ is computed from the CDF of the Weibull clutter distribution as shown in Fig.~\ref{Fig:Adaptivethreshold}.

\begin{figure}[!t]
 \centering\vspace{-3mm}
 \includegraphics[width=0.42\textwidth]{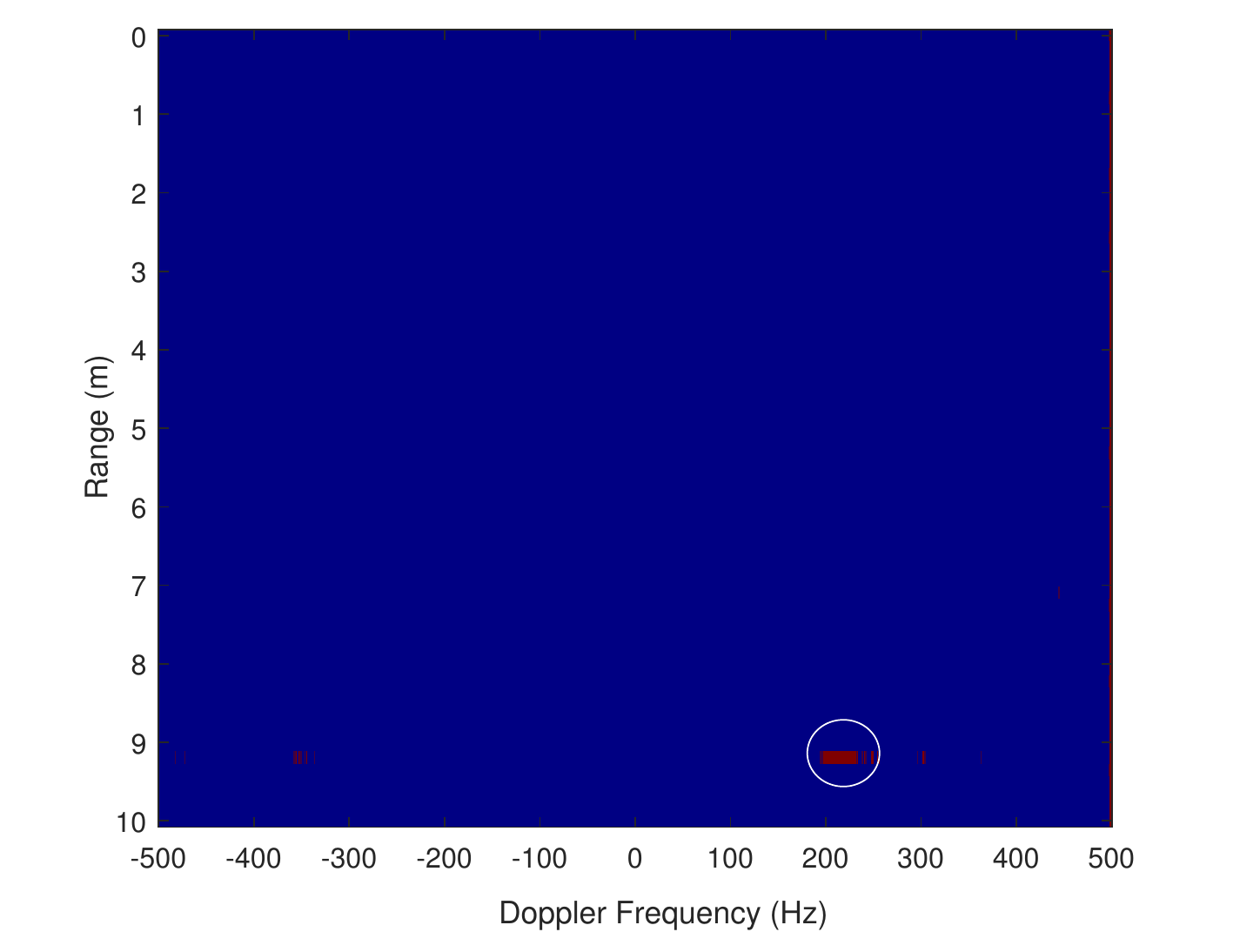}\vspace{-1mm}
 \caption{Experimental UAV detection on the range-Doppler map with $P_{\rm fa}=10^{-5}$ and $N=20$. The experiment is carried out using a 25 GHz Ancortek radar.} \label{Fig:CFAR_Detection}\vspace{-5mm}
\end{figure}

Fig.~\ref{Fig:CFAR_Detection} shows the experimental result of the CFAR detection algorithm. The target is a low altitude micro-UAV (DJI Phantom4 Pro) flying in a rocky terrain. The clutter power is modeled using the Weibull distribution. The UAV is detected using a 25 GHz mmWave Ancortek radar with properties shown in Table~\ref{Table:Table1}. For a given mmwave radar with transmit frequency ($f_{o}$), the range and Doppler frequency shift ($f_{Doppler}$) of the target UAV are obtained directly from the range-Doppler plot. The target velocity is estimated from the $f_{Doppler}$ using $v\approx \frac{f_{Doppler}*c}{2*f_{o}}$. From the range-Doppler plot, we see the range is about 9 meters and $f_{Doppler}$ is 200 Hz. The estimated velocity is 1.2 meters/seconds. In our experiment, the micro-UAV was moving slowly.

% Further discrimination of the UAV is achieved by extracting the micro-Doppler signatures of its rotating propellers as shown in Fig.~\ref{Fig:RangeDoppler}~(right).

% The adaptive threshold is described further in Fig.~\ref{Fig:adaptive_threshold1}. \textcolor{red}{IG: It is not clear what the figure shows. Or it is obvious, I think better to remove Fig. 3.}
 \nonumber

\begin{figure}[t!]
 \center\vspace{-3mm}
 \includegraphics[width=0.42\textwidth]{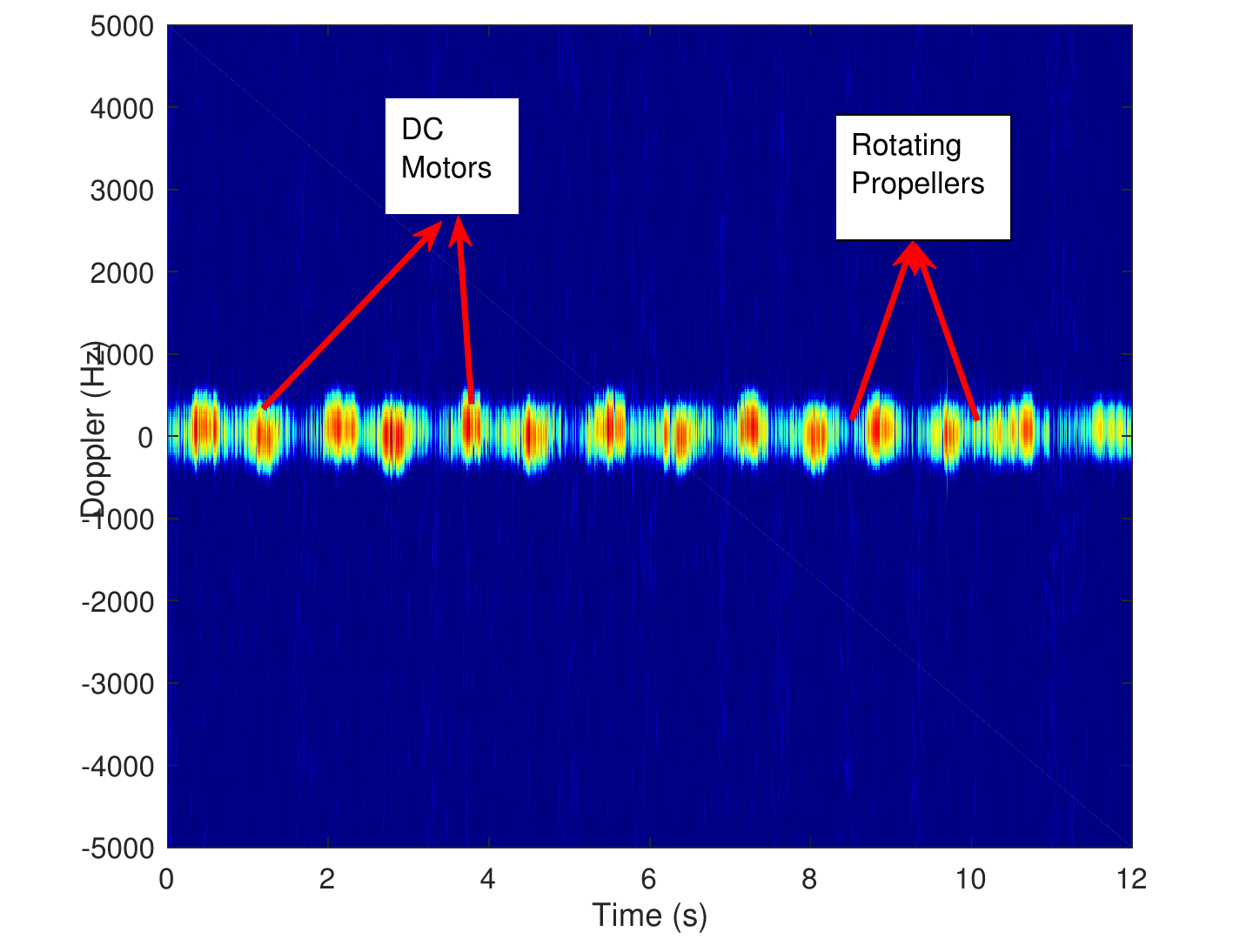}\vspace{-1mm}
 \caption{Experimental measurement of the radar micro-Doppler signature of the DJI Phantom 4 Pro quadcopter UAV using the 25 GHz Ancortek radar.}\label{Micro-Doppler_UAV3.eps}\vspace{-3mm}
\end{figure}

\begin{figure}[t!]
 \center\vspace{-1mm}
 \includegraphics[width=0.42\textwidth]{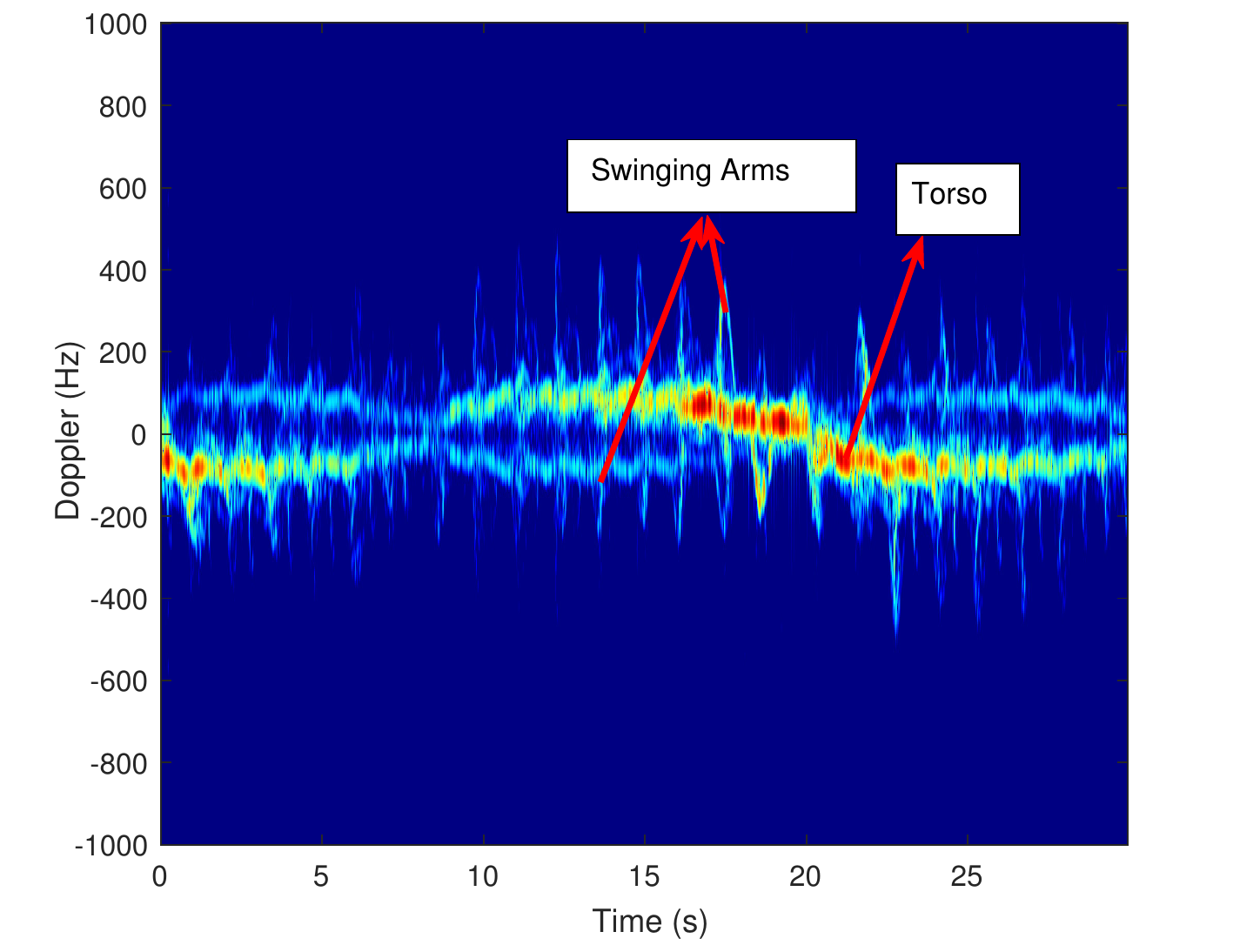}\vspace{-1mm}
 \caption{Experimental measurement of the radar micro-Doppler signature of a walking man with swinging arms using the 25 GHz Ancortek radar.  }\label{Fig:Walking_man}\vspace{-3mm}
\end{figure}

\begin{figure}[t!]
 \center\vspace{-1mm}
 \includegraphics[width=0.42\textwidth]{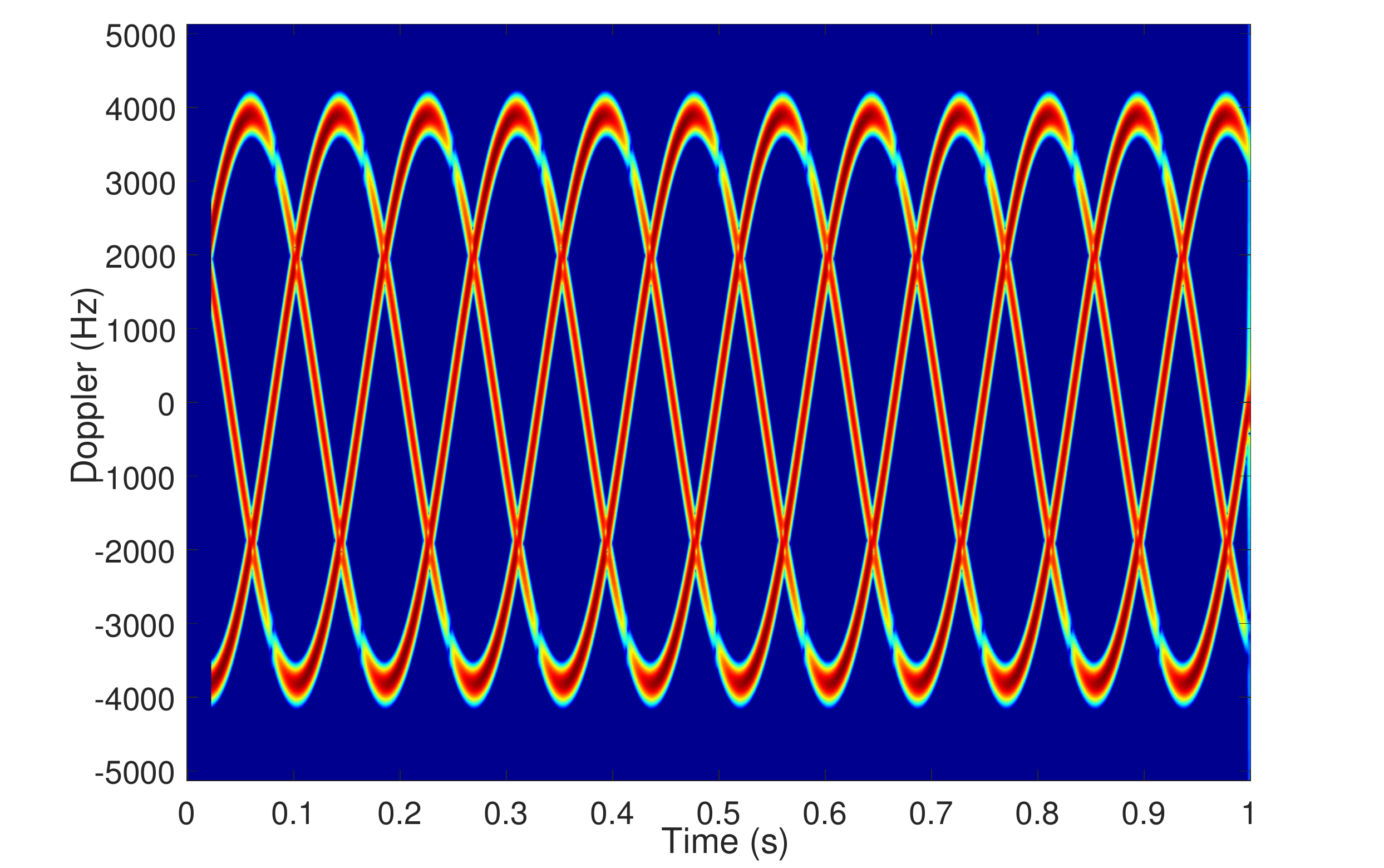}\vspace{-1mm}
 \caption{Simulated micro-Doppler signature of a rotating three-bladed toy helicopter~\cite{chen2014radar}.}\label{Fig:heicopter}\vspace{-3mm}
\end{figure}

% \begin{align}
% %\label{eq:12}
% f_{\rm md} = \frac{2f\Omega}{c}\left[\hat{\omega}'^2\sin{\Omega t}\nonumber-\hat{\omega}'^3\cos{\Omega t} + \hat{\omega}'(\mathbf{I}+\hat{\omega}'^2){\mathfrak{R}_{\rm I}}\cdot\vec{r}_0\right]^T \cdot\textbf{n},
% \nonumber
% \end{align}

\subsection{The UAV micro-Doppler Signature}
 
Micro-UAVs have at least one pair of fast moving propellers which provide the required aerodynamic lift and spatial maneuverability for the UAV. The radar backscattered signal from these rotating propellers modulate the main Doppler frequency shift induced by the bulk translational motion of the UAV. These modulation are referred as the micro-Doppler signatures of the UAV. Micro-Doppler signatures can help discriminate UAVs from other airborne objects such as birds, kites and commercial aircrafts. According to~\cite{chen2014radar}, the micro-Doppler frequency shift induced by a rotating blade is described by:
%\begin{eqnarray}
%\label{eq:12}
$f_{\mu{\rm D}} =\frac{2f\Omega}{c}[\boldsymbol{\hat{\omega}'}^2\sin{\Omega t}-\boldsymbol{\hat{\omega}'}^3\cos{\Omega t}%\\ 
\nonumber+\boldsymbol{\hat{\omega}'}(\mathbf{I}+\hat{\omega}'^2)\mathfrak{R}_{\rm Init}\cdot\vec{r}_0]$,
%\end{eqnarray}
where $f$ is the operational frequency of the radar, $\Omega$ is the scalar angular velocity of the rotating blade (propeller), $c$ is the speed of light, $\vec{r}_0$ is the initial position vector of the propeller, $\mathbf{I}$ is the identity matrix, $\mathfrak{R}_{\rm I}$ is the initial rotation matrix of the propeller defined by the Euler-Rodrigues rotation theorem, $\boldsymbol {\hat{\omega}'}$ is a skew symmetric matrix associated with the unit vector in the direction of the angular rotation velocity of the propeller, $\mathbf{\omega}$, and \textbf{n} is the unit vector of the radar line-of-sight (LOS). Accordingly, %to (\ref{eq:12}), 
the micro-Doppler frequency shift %, $f_{\rm md}$, 
induced by the UAV propeller is a time varying function whose effect becomes more significant when a high frequency radar is used. % to illuminate the UAV. 

 Fig.~\ref{Micro-Doppler_UAV3.eps} shows the experimental micro-Doppler signature of the DJI Phantom 4 Pro quadcopter UAV. The features of this UAV (metallic DC motor and rotating propeller blades) can be easily recognized. For comparison, the experimental micro-Doppler signature of a walking man and simulated micro-Doppler signature of a three bladed toy helicopters~\cite{chen2014radar} are also generated in Fig.~\ref{Fig:Walking_man} and Fig.~\ref{Fig:heicopter} respectively. Comparing these figures, it is obvious that the micro-UAV can easily be discriminated from the other objects by its unique micro-Doppler signature.

\section{Numerical Results and Conclusion}\label{Sec:NumRes}

\begin{figure}[!t]
\vspace{-0.15cm}
 \center
 \includegraphics[width=0.42\textwidth]{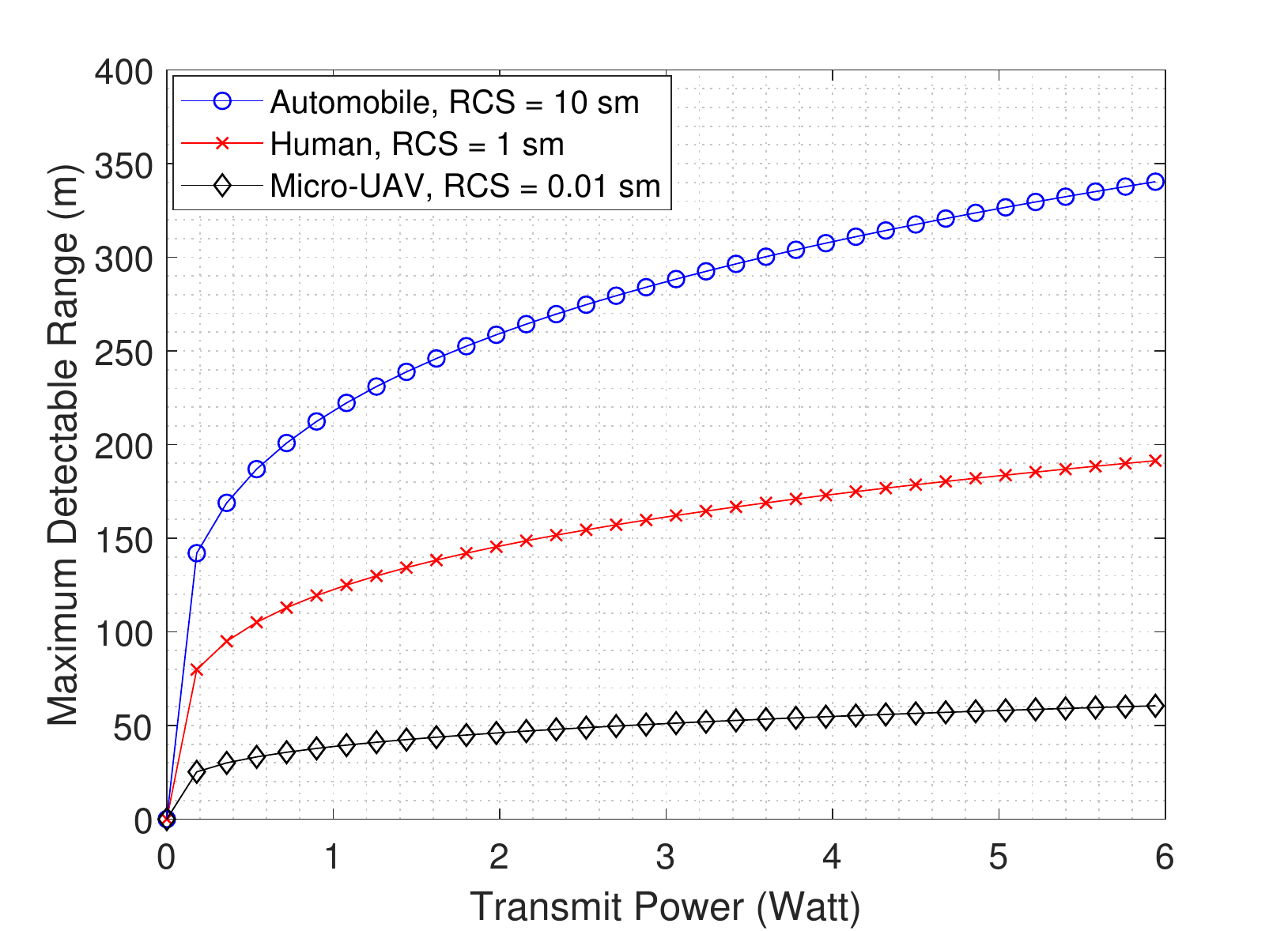}
 \caption{Radar maximum range for three different targets.}\vspace{-3mm}\label{Fig:MaxRangePlot}\vspace{-0.1cm}
\end{figure}

\begin{figure}[t!]
\vspace{-0.1cm}
 \center
 \includegraphics[width=0.42\textwidth]{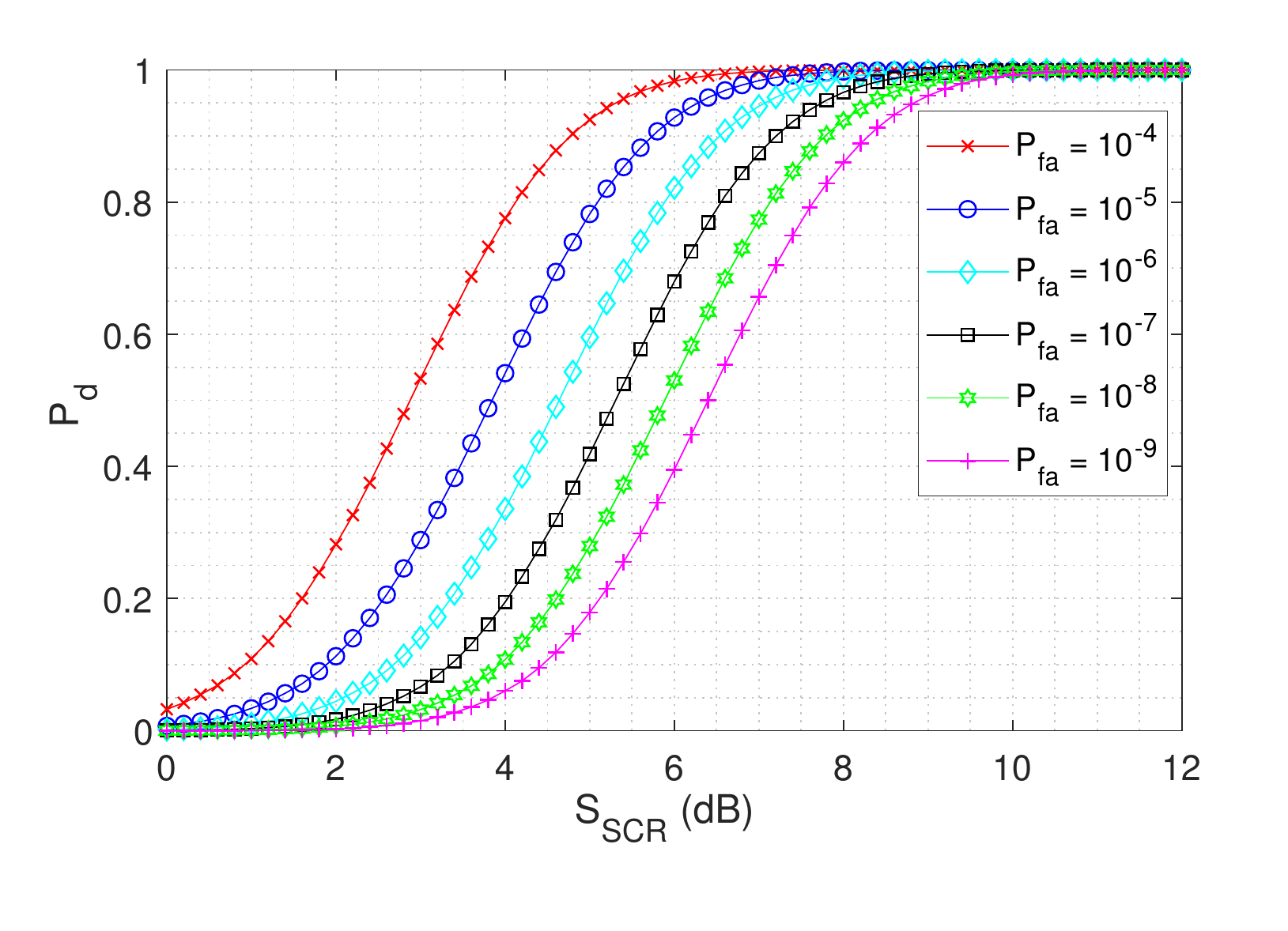}\vspace{-6mm}
 \caption{ROC curve showing simulated $P_{\rm d}$ versus SCR.}\vspace{-3mm}\label{Fig:ROC_Curve}
 
\end{figure}

\begin{figure}[t!]
\vspace{-0.15cm}
 \center
 \includegraphics[width=0.42\textwidth]{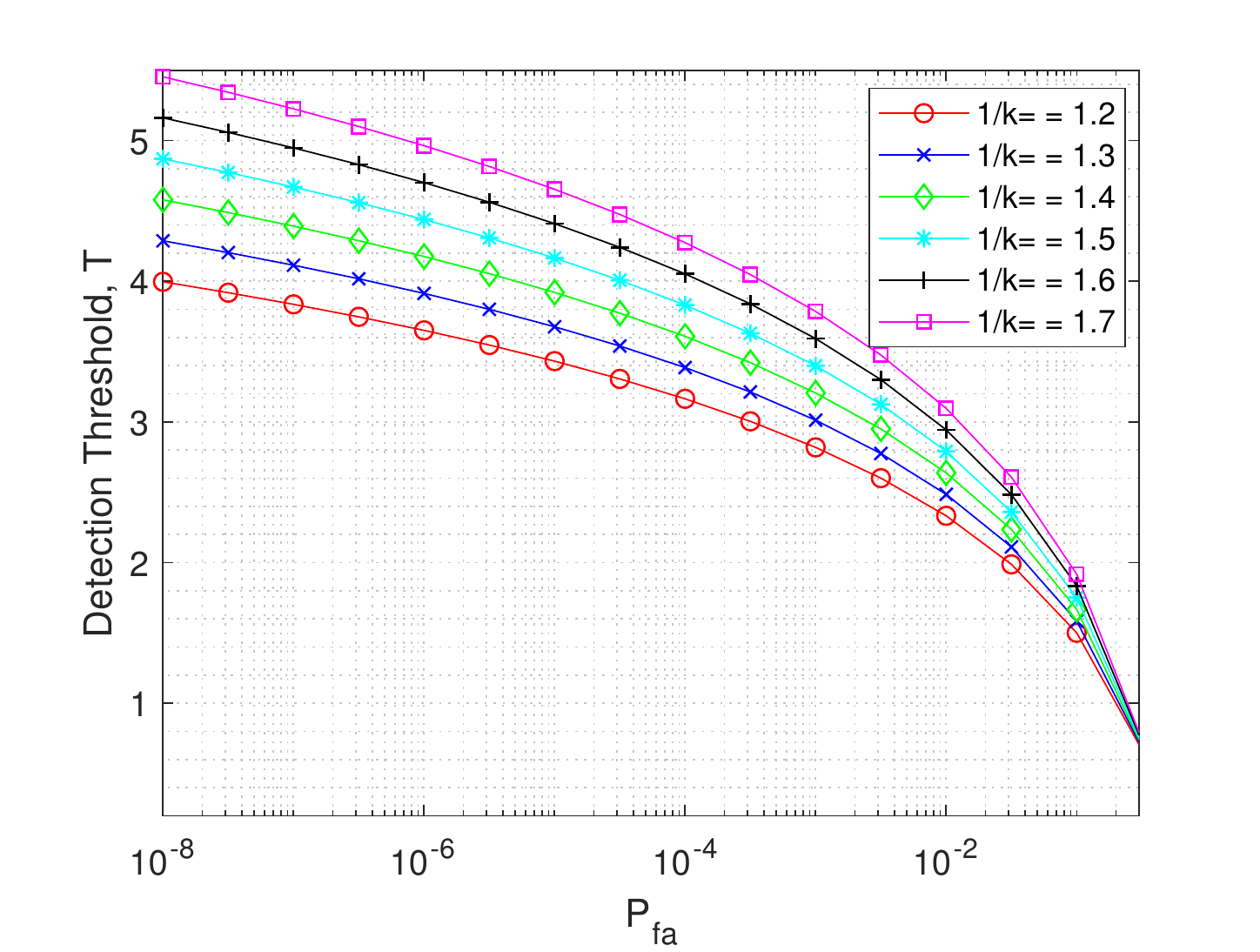}\vspace{-3mm}
\caption{Detection threshold versus $P_{\rm fa}$ for $\hat{b}_{\rm MLE}=1.65$.}\vspace{-5mm}\label{Fig:DetectionThreshodvsPfa}
\end{figure}

Figs.~\ref{Fig:MaxRangePlot}-\ref{Fig:DetectionThreshodvsPfa} provide simulation result for the detection of low altitude micro-UAV using a 25 GHz mmWave radar. Fig.~\ref{Fig:MaxRangePlot} shows that the maximum detectable range of the radar depends on the RCS of the micro-UAV (expressed in square meters (sm)) and the transmit power of the radar. In Fig.~\ref{Fig:ROC_Curve} and Fig.~\ref{Fig:DetectionThreshodvsPfa}, we see the effect of the land clutter distribution on the detection performance of the CFAR algorithm. It is obvious from Fig.~\ref{Fig:ROC_Curve}, that the $P_{\rm d}$ of the micro-UAV increases with $S_{\rm SCR}$. Therefore, for a desired $P_{\rm fa}$, the $P_{\rm d}$ can be improved by increasing the $S_{\rm SCR}$. 

Fig.~\ref{Fig:DetectionThreshodvsPfa} shows that for a desired $P_{\rm fa}$, the optimal adaptive threshold, $T$, of the CFAR algorithm will depend on the properties of the land clutter distribution: the shape ($k$) and scale ($b$) parameters. This is because a radar land clutter distribution will have different values of $k$ and $b$ at different spatial regions. From \eqref{eq:11} % equation ~\ref{eq:11}, 
we can see that $P_{\rm fa}\approx \exp[-[\frac{T}{b}]^k]$. If the original threshold $T$ is set at 10 dB and $\hat{b}_{\rm MLE}$ is estimated as 1.67. For a small change in the $k$ value of the clutter distribution, say from 2.0 to 1.5, there will be more than $10^{6}$ fold increment in $P_{\rm fa}$ given that $\hat{b}_{\rm MLE}$ stays almost the same. Therefore, the CFAR algorithm is required to adaptively change the threshold for different operational domain with different clutter distribution. In Fig.~\ref{Fig:DetectionThreshodvsPfa}, we see how the CFAR detector will alter $T$ for different clutter properties. In practice, the values of $k$ and $b$ for a given operational region is estimated from actual clutter data. The clutter data can be modeled with Weibull distribution to provide an improved target detection performance.


\section{Conclusion}

%The experimental and simulation 
Our results show how %that 
the optimal detection performance of a low-grazing angle mmWave radar depends on the RCS of the target micro-UAV, radar properties, and the properties of the land clutter. In addition, we experimentally verified that the micro-Doppler signature of the target micro-UAV is unique. This information can be used to discriminate the target from other objects.

% \begin{figure*}[t!]
%     \centering
%     \begin{subfigure} %{0.4\textwidth} 
%     % width of left subfigure
%         \centering
%       \includegraphics[width=0.25\textwidth]{Max_Range_plot.pdf}
%         \caption{Lorem ipsum}
%     \end{subfigure}%
%  % add desired spacing between images, e. g. ~,\vspace{1em}, \quad, \qquad etc.
% %(or a blank line to force the subfigure onto a new line)
%     \begin{subfigure}[t] %{0.5\textwidth}
%         \centering
%         \includegraphics[width=0.25\textwidth]{Max_Range_plot.pdf}
%         \caption{Lorem}
%     \end{subfigure}
%     \caption{Caption place holder}
% \end{figure*}

% \begin{figure}[!h]
% 	\centering
% 	\begin{minipage}[t]{2cm}
% 		\centering
% 		\includegraphics[scale=0.3]{Max_Range_plot.pdf}
% 		\caption{Pièce (1) - Fonte}
% 	\end{minipage}
% 	\hspace{2cm}
% 	\begin{minipage}[t]{2cm}
% 		\centering
% 		\includegraphics[scale=0.3]{Max_Range_plot.pdf}
% 		\caption{Pièce (2) - Aluminium}
% 	\end{minipage}
 
% 	\begin{minipage}[t]{2cm}
% 		\centering
% 		\includegraphics[scale=0.3]{Max_Range_plot.pdf}
% 		\caption{Caméra thermique 1}
% 	\end{minipage}
% 	\hspace{2.5cm}
% 	\begin{minipage}[t]{2cm}
% 		\centering
% 		\includegraphics[scale=0.3]{Max_Range_plot.pdf}
% 		\caption{Caméra thermique 2}
% 	\end{minipage}
% \end{figure}

\bibliographystyle{IEEEtran}
\bibliography{Ref_APL}

% \clearpage 

 \end{document}